\documentclass[12pt]{article}
\usepackage{amssymb}
\usepackage{amsmath}
\usepackage{graphicx,color}

\usepackage{graphics}
\usepackage{color}
\usepackage{amssymb,amsmath}
\usepackage{slashed}
\usepackage{epstopdf}
\usepackage{booktabs}
\usepackage{mathrsfs}
\usepackage[export]{adjustbox}
\usepackage{epsfig}
\usepackage{graphicx,color}

\newcommand{\bea}{\begin{eqnarray}}
\newcommand{\eea}{\end{eqnarray}}
\newcommand{\sy}{s(y)}

\numberwithin{equation}{section}

\begin{document}
%
\begin{titlepage}
\vspace*{10mm}
\begin{center}
\baselineskip 25pt 
{\Large\bf
Domain-Wall Standard Model in non-compact 5D \\ and LHC phenomenology
}
\end{center}
\vspace{5mm}
\begin{center}
{\large
Nobuchika Okada, 
Digesh Raut, and  
Desmond Villalba
}

\vspace{.5cm}

{\baselineskip 20pt \it
Department of Physics and Astronomy, \\
The University of Alabama, Tuscaloosa, AL35487, USA\\
} 

\end{center}
\vspace{0.5cm}
\begin{abstract}
We propose a framework to construct ``Domain-Wall Standard Model'' in a non-compact 5-dimensional space-time, 
  where all the Standard Model (SM) fields are localized 
  in certain domains of the 5th dimension and the SM is realized as a 4-dimensional effective theory 
  without any compactification for the 5th dimension. 
In this context, we investigate the collider phenomenology of 
  the Kaluza-Klein (KK) modes of the SM gauge bosons and the current constraints 
  from the search for a new gauge boson resonance at the Large Hadron Collider Run-2. 
The couplings of the SM fermions with the KK-mode gauge bosons depend 
  on the configuration of the SM fermions in the 5-dimensional bulk. 
This ``geometry'' of the model can be tested 
  at the future Large Hadron Collider experiment, once a KK-mode of the SM gauge boson is discovered. 

\end{abstract}
\end{titlepage}

\section{Introduction}

A possibility that our world consists of more than 4-dimensional space-time
  has been attracting a good deal of attention for a long time. 
After the discovery of the D-brane in string theories \cite{D-brane}, the brane-world scenario 
  has been intensively studied as new physics beyond the Standard Model (SM). 
In extra-dimensional theories, a new property, ``geometry,'' comes into play in phenomenology 
  and provides us with a new possibility of understanding mysteries within the SM. 
The large extra-dimension model \cite{ADD} is a well-known brane-world scenario 
  which offers a solution to the gauge hierarchy problem by ``diluting" the Planck scale 
  to the TeV scale by a large extra-dimensional volume. 
Another well-known scenario is the warped extra-dimension model \cite{RS}, 
  where the Planck scale is ``warped down" to the TeV scale 
  in the presence of the anti-de Sitter (AdS) curvature in the extra 5th dimension.

In most of the extra-dimensional models that have been investigated until now, 
  extra-dimensions are assumed to be compactified on some manifolds 
  or orbifolds, and they are treated differently from our 3-spatial dimensions. 
We may suppose it more natural if all spatial dimensions are non-compact 
  and the SM  is realized as a 4-dimensional effective theory. 
This picture requires that all SM fields as well as 4-dimensional graviton
  are localized in certain 3-spatial dimensional domains in the bulk space. 
A simple realization of this picture for 4-dimensional graviton 
  has been proposed in Ref.~\cite{RS2}, the so-called RS-2 scenario, 
  where due to the 5-dimensional AdS curvature, 
  4-dimensional graviton is localized at a point in non-compact 5th dimension 
  and the Einstein gravity in 4 dimensions is reproduced at low energies.

In this paper, we propose a framework to construct ``Domain-Wall Standard Model" in 5-dimensional space-time, 
  where all the SM fields, including the gauge bosons, are localized in certain domains along the 5th dimension. 
This direction has been considered in Ref.~\cite{Davies:2007xr} before, where the SU(5) gauge symmetry is introduced 
  in the 5-dimensional bulk and the Dvali-Shifman mechanism \cite{Dvali:1996xe} is assumed for dynamically localizing 
  the SM gauge bosons associated with the breaking of the SU(5) gauge symmetry down to the SM gauge group. 
In this paper, we consider a simple way for localizing a gauge field in the bulk without extending the SM gauge groups, 
  which has been proposed in Ref.~\cite{OS}.  
We utilize the same mechanism for the localization of the SM Higgs field, 
  while the SM fermions are localized via the mechanism proposed in Ref.~\cite{rubakov} (domain-wall fermion).  
Our scenario is a field-theoretical realization of a ``3-brane" in which all the SM fields are confined. 
However, the finite width of the ``3-brane" leads to rich physics phenomenologies  
 that can be explored in future Large Hadron Collider (LHC) experiments.

In the next section, we describe the basic construction of the Domain-Wall SM, where all the SM fields   
  (the gauge bosons, the Higgs field, and the chiral fermions) are localized via certain mechanisms. 
Effective couplings between Kaluza-Klein (KK) modes of the SM gauge bosons and 
  the SM fermions are derived in the 4-dimensional effective theory. 
We discuss the collider phenomenology on KK-modes of the SM gauge bosons in Sec.~\ref{sec:3}. 
The current results from the search for a new gauge boson resonance at the LHC Run-2 
  is interpreted as constraints on our model.    
In the last section, we summarize our results and discuss future directions for the Domain-Wall SM.

\section{Basic construction of Domain-Wall Standard Model}
\label{sec:2}

In realizing the Domain-Wall SM, we need localization mechanisms 
 for the myriad of SM fields: the gauge fields,  the Higgs field, and the chiral fermions. 
In this section, we present the basics construction of the Domain-Wall SM. 

\subsection{Gauge field localization} 
Since the essence for localizing a gauge field is independent of the gauge structure, 
  we address the gauge field localization based on a U(1) gauge theory. 
A simple way of localizing the gauge field has been proposed in Ref.~\cite{OS}, 
   where a gauge coupling depending on the 5th coordinate is introduced.\footnote{
The same idea was discussed elsewhere before Ref.~\cite{OS}. 
See, for example, Ref.~\cite{LO}     
}    
In 5-dimensional flat Minkowski space-time, the basic Lagrangian for the U(1) gauge field is of the form,
\bea
\mathcal{L}_{5}=-\frac{1}{4}s(y)F_{MN}F^{MN}, 
\eea
where $F_{MN}$ is the gauge field strength, and $M, N=0,1,2,3,y$ with $y$ being the index for the 5th extra-dimensional coordinate. 
Our convention for the metric is $\eta_{MN}={\rm diag}(1,-1,-1,-1,-1)$, and in general we suppress coordinate dependence 
  of the fields unless emphasis is needed. 
Note that the gauge coupling $1/g^2=s(y)$ depends on the 5th coordinate $y$.  
Decomposing the field strength into its components yields the following expression (up to total derivative terms):
\bea
\mathcal{L}_{5}
 &=&
 \frac{1}{2} s A^{\mu}\left(g_{\mu \nu}\Box_{4} -\partial_{\mu}\partial_{\nu}\right)A^{\nu}-\frac{1}{2} s A_{y}\Box_{4}A_{y} 
\nonumber \\
 &-& \frac{1}{2}A_{\mu}\partial_{y}\left(s \partial_{y}A^{\mu}\right)-\left(\partial_{\mu}A^{\mu}\right)\partial_{y}\left(s A_{y}\right),
\label{L5} 
\eea
where $A_{\mu}$ $(\mu=0,1,2,3)$ and $A_y$ are a gauge field and a scalar field 
  in 4-dimensional space-time, and the first line in the right-hand-side denotes the 4-dimensional kinetic terms for these fields.\footnote{
In literature, authors sometimes employ the ``axial gauge'' $A_y=0$ to simplify the 5-dimensional equations of motion for the gauge field. 
However, this procedure may be misleading, since as we will see in the following, the zero-mode of $A_y$  vanishes not by the gauge-fixing but due to the explicit breaking of the 5-dimensional gauge invariance by $s(y)$. 
This is analogous to the 5-dimensional models with a $S_1$/$Z_2$ orbifold compactification, where 
$A_y$ for the zero-mode is projected out by the orbifolding, but not by the axial gauge-fixing.
In our discussion, the gauge transformation is employed to eliminate 2 degrees of freedom from $A_\mu$ 
  as usual in 4-dimensional gauge field theories, and  $A_y$ is left as a dynamical field.   
}   
The last term contains a mixing between $A_{\mu}$ and $A_y$. 
Note that this structure is analogous to that in spontaneously broken gauge theories, 
  and we eliminate the mixing term by adding a gauge fixing term, which is a 5-dimensional analog to the $R_{\xi}$ gauge:\footnote{
During the completion of our manuscript, we have learned that another group is considering a more general procedure for localized gauge fields \cite{Arai:2018rwf}.
}   
\bea
\mathcal{L}_{\rm GF}=-\frac{s}{2 \, \xi}\left(\partial_{\mu}A^{\mu}- \frac{\xi}{s} \partial_{y}(s A_y)\right)^2 ,
\eea
where $\xi$ is a constant parameter. 
The total Lagrangian now reads $\mathcal{L}=\mathcal{L}_{5}+\mathcal{L}_{\rm GF}=\mathcal{L}_{gauge}+\mathcal{L}_{scalar}$, where
\bea 
\mathcal{L}_{gauge}&=&\frac{1}{2} s A^{\mu}\left(g_{\mu \nu}\Box_{4}-\left(1-\frac{1}{\xi}\right)\partial_{\mu}\partial_{\nu}\right)A^{\nu}-\frac{1}{2}A_{\mu}\partial_{y}(s \, \partial_{y}A^{\mu}), 
\label{gauge}
\\ 
\mathcal{L}_{scalar}&=&-\frac{1}{2} s  A_{y}\Box_{4} A_{y}+\frac{1}{2} s \, \xi A_{y}\partial_{y} \left(\frac{1}{s}\partial_{y}(s A_{y}) \right).
\label{scalar}
\eea

Next, we analyze the Kaluza-Klein (KK) modes of the gauge and scalar fields via the mode expansions
\bea
A_{\mu}(x,y)=\sum _{n=0}^\infty A_{\mu}^{(n)}(x)\chi^{(n)}(y), \quad A_{y}(x,y)=\sum _{n=0}^\infty \eta ^{(n)}(x)\psi^{(n)}(y),  
\eea
where $x=x^\mu$. 
From Eqs.~(\ref{gauge}) and (\ref{scalar}), we obtain the KK-mode equations: 
\bea
\frac{d}{dy}
\left(s \,  \frac{d}{dy}\chi^{(n)} \right)+ s \, m_{n}^{2}\chi^{(n)}=0,  \; \; \; 
\frac{d}{dy}\left( \frac{1}{s} \frac{d}{dy} (s \psi^{(n)} ) \right)+ \tilde{m}_{n}^{2}\psi^{(n)}=0. 
\label{KK_EOM}
\eea
With the solutions of these KK-mode equations, the Lagrangians in Eqs.~(\ref{gauge}) and (\ref{scalar}) are written as 
\bea 
\mathcal{L}_{gauge}&=& \sum_{n=0}^\infty \frac{1}{2} s  \left( \chi^{(n)} \right)^2
 \left[ A_{\mu}^{(n)} \left(g^{\mu \nu}(\Box_{4} + m_n^2) -\left(1-\frac{1}{\xi}\right)\partial^{\mu}\partial^{\nu}\right)A_{\nu}^{(n)}  \right] , 
 \nonumber \\ 
\mathcal{L}_{scalar}&=&- \sum_{n=0}^\infty  \frac{1}{2} s   \left( \psi^{(n)} \right)^2  
  \left[ \eta^{(n)} \left( \Box_{4} + \xi \, \tilde{m}_n^2 \right) \eta^{(n)} \right], 
\label{L2} 
\eea
%
Note that if the two equations in Eq.~(\ref{KK_EOM}) have solutions 
  with $m_n=\tilde{m}_n$, Eq.~(\ref{L2}) indicates that 
  the relation between $A_\mu^{(n)}$ and $\eta^{(n)}$ in the 4-dimensional effective theory 
  is nothing but the one between a massive gauge boson and a would-be Nambu-Goldstone (NG) mode 
  in the $R_{\xi}$ gauge after the spontaneous gauge symmetry breaking.  
In this case, we can identify the KK-modes of $\eta^{(n)}$ with would-be NG modes  
  eaten by the KK-modes of $A_\mu^{(n)}$.  
From the view point of 4-dimensional effective theory, this picture seems quite reasonable.   
In fact, we find $m_n=\tilde{m}_n$ for the solvable example that we discuss in the following.

Even for a general function of $s(y)$, it is easy to find zero-mode solutions with $m_0=0$ in Eq.~(\ref{KK_EOM}). 
General solutions are given by 
\bea
\chi^{(0)}=\tilde{c}_{\chi}+c_{\chi}\int^{y} dy'\frac{1}{s(y')}, \; \; \; 
\psi^{(0)}=\frac{\tilde{c}_{\psi}}{\sy}+\frac{c_{\psi}}{\sy}\int^{y} dy's(y'),
\eea
where $ \tilde{c}_{\chi}$, $c_{\chi}$, $\tilde{c}_{\psi}$, and $c_{\psi}$ are constants. 
In order to localize the gauge field in the finite domain,  we impose $\sy \to 0$ as $|y| \to \infty$. 
In addition, the gauge field in the 4-dimensional effective theory must be normalizable in the sense that
\bea\label{norm}
\int_{-\infty}^{\infty} dy \, \sy \, \chi^{(n)}(y) \, \chi^{(n)} (y) < \infty.
\eea
Considering the zero-mode solution for the gauge field, this constraint requires $c_{\chi}=0$, 
   resulting in the zero-mode for the gauge boson having a constant configuration in the 5th coordinate direction. 
Note that this forces the gauge coupling to be universal in the 4-dimensional effective theory 
  and hence the 4-dimensional gauge invariance is maintained. 
Applying the same logic to the scalar $A_y$ component yields an interesting result: 
 the solution of $\psi^{(0)}$ cannot satisfy the requirement given in Eq.~(\ref{norm}) unless $c_{\psi}=\tilde{c}_{\psi}=0$. 
This suggests to us that $\psi^{(0)}=0$ is the only appropriate choice for the zero-mode of the scalar. 
Hence, no (normalizable) zero-mode exists for the scalar component. 
Note that if $s(y)$ is independent of $y$, a constant $\psi^{(0)}$ becomes a consistent solution. 
This is a trivial case where the 5-dimensional gauge invariance is manifest. 
Therefore, the absence of the zero-mode scalar originates from the explicit breaking of the 5-dimensional 
  gauge invariance due to $y$-dependence of the gauge coupling $s(y)$.

For the following discussion about the LHC phenomenology, let us consider a solvable example for $s(y)$ as 
\bea
 \sy =
  \begin{cases}
    S_0 + \epsilon      & \quad (|y|< L)\\
    \epsilon & \quad (|y|> L)
  \end{cases}
  =S_0 \left[ H(y+ L)-H(y-L) \right]+\epsilon, 
\eea
where $S_0$ and $\epsilon$ are real, positive constants,  
  $H(x)$ is the Heaviside function, 
  and we have introduced the small parameter $\epsilon \ll S_0$ to regularize $1/s(y)$. 
Since $s(y)$ is invariant under the reflection of the 5th coordinate $y \to -y$, 
  let us simplify our system by introducing $Z_2$-parity. 
Under $y \to -y$, the Lagrangian in Eq.~(\ref{L5}), and $A_\mu$ are even, while $A_y$ is odd. 
Now we can easily find the solution to the KK-mode equations in Eq.~(\ref{KK_EOM}) for $y \neq \pm L$ 
  and the KK-mode expansions are expressed as 
\bea
A_{\mu}(x,y)= A_\mu^{(0)}(x) + \sum_{n=1}^\infty A_{\mu}^{(n)}(x) \cos(m_n y), 
  \quad 
A_{y}(x,y)=\sum_{n=1}^\infty \eta ^{(n)}(x) \sin(m_n y). 
\eea
Although the zero-mode of $A_y$ is projected out because of the $Z_2$-parity in the present system,  
   the zero-mode doesn't exist in any cases as we have discussed above. 
Since $d s(y)/dy$ is singular at $y= \pm L$, boundary conditions are imposed to make the solutions regular, 
  namely, 
\bea
  \frac{d}{dy}A_\mu(x, y) \left|_{y\to \pm L}  \right. =0,  \quad  A_y(x,y)\left|_{y\to \pm L} \right. =0. 
\eea  
Thus we find the mass eigenvalue for each KK-mode of $A_\mu^{(n)}$ as $m_n= n (\pi/L)$, 
    while the mass for each KK-mode of $\eta^{(n)}$, which is a would-be NG mode 
    eaten by the corresponding $A_\mu^{(n)}$, is given by $\xi m_n^2$. 
By integrating out the 5th-dimensional degrees of freedom and taking a limit $\epsilon \to 0$,  
    the 4-dimensional effective Lagrangian is then
\bea
\mathcal{L}_4&=&\int_{-\infty}^\infty dy \, \mathcal{L}_5 \nonumber \\ 
&=& \frac{1}{2} \left( \frac{1}{g} \right)^2 A_{\mu}^{(0)}\left[g^{\mu \nu}\Box_{4}-\left(1-\frac{1}{\xi}\right)\partial^{\mu}\partial^{\nu}\right]A_{\nu}^{(0)} \nonumber \\
&+&\frac{1}{2} \left( \frac{1}{\sqrt{2} \, g} \right)^2 
\sum_{n=1}^{\infty} \left\lbrace A_{\mu}^{(n)}\left[g^{\mu \nu}\left(\Box_{4}+m_{n}^2\right)
-\left(1-\frac{1}{\xi}\right)\partial^{\mu}\partial^{\nu}\right]A_{\nu}^{(n)}\right\rbrace \nonumber \\
&-&\frac{1}{2} \left( \frac{1}{\sqrt{2} \, g} \right)^2  \sum_{n=1}^{\infty} \left[ \eta^{(n)} \left(\Box_{4}+\xi m_{n}^2\right) \eta^{(n)} \right], 
\label{4DA}
\eea
where we have defined the gauge coupling  ($g$) in the 4-dimensional effective theory as 
  $1/g^2=2 S_0 L$. 
Note that this effective Lagrangian is the same as the one obtained from a 5-dimensional gauge theory 
  by compactifying the 5th dimension on $S^1/Z_2$ orbifold with a radius $L/\pi$, (see, for example, Ref.~\cite{UED}). 
For this solvable example, we have obtained an effectively compactified 5-dimensional gauge theory. 
However, as we will see in the following, the KK-model phenomenology is quite different, 
  since the SM fermions have non-trivial configurations in the 5th dimension.

\subsection{Localized Higgs field} 
Next we consider the 5-dimensional extension of the Higgs mechanism. 
To simplify our discussion, we take the Abelian Higgs model as an example, 
  corresponding to our previous discussion about the localized U(1) gauge field.  
It is straightforward to extend our discussion to the SM Higgs doublet case.

In non-compact extra-dimensions, we need to consider a localization mechanism 
 for not only Higgs field but also its vacuum expectation value. 
For this purpose, we may utilize the same procedure taken for the gauge field, 
 and the Lagrangian for the Higgs sector is defined as 
\bea
\mathcal{L}_5^H=\sy \left[ ({\cal D}^{M}H)^{\dagger}({\cal D}_{M}H)
  -\frac{1}{2}\lambda \left(H^{\dagger}H-\frac{v^2}{2} \right)^2 \right], 
\label{L_H}  
\eea
where $H$ is the Higgs field, $\lambda$ is a Higgs quartic coupling,
  $v$ is its vacuum expectation value, 
  and the covariant derivative is given by ${\cal D}_M= \partial_M - i Q_H A_M$ 
  with a U(1) charge $Q_H$ for the Higgs field.   
Here, we define the Higgs field as a $Z_2$-even field with a mass dimension $1$.

Expanding about the vacuum $H=(v + h + i \phi)/\sqrt{2}$ and neglecting the interaction terms, 
  we obtain (up to total derivative terms)
\bea
\mathcal{L}_5^H & \supset & \frac{1}{2} \sy \left[ (\partial^{M}h)(\partial_{M}h)-m_{h}^{2}h^2\right]
 +\frac{1}{2}\sy(\partial^{M}\phi)(\partial_{M}\phi) \nonumber \\
&=&-\frac{1}{2} s h(\Box_{4}+m_{h}^{2})h+\frac{1}{2}h\,\partial_y(s \partial_{y}h) 
  -\frac{1}{2} s \phi\Box_{4}\phi+\frac{1}{2}\phi\,\partial_{y}(s \partial_{y}\phi) , 
\eea
where $m_h^2= \lambda v^2$ is the physical Higgs boson mass. 
From the KK-mode decomposition for these fields, 
\bea
h(x,y)=\sum_{n=0}^{\infty}h^{(n)}(x)\chi^{(n)}_{h}(y), 
  \quad 
\phi(x,y)=\sum_{n=0}^{\infty}\phi^{(n)}(x)\chi^{(n)}_{\phi}(y), 
\eea
   we can see that the KK-mode equations for $\chi^{(n)}_{h}$ and $\chi^{(n)}_{\phi}$ 
   are identical to that of the gauge boson in Eq.~(\ref{KK_EOM}). 
Since the zero-mode $\phi^{(0)}$ is the would-be NG mode eaten 
   by $A_\mu^{(0)}$, the theoretical consistency requires the configurations of $\phi^{(0)}$ and $A_\mu^{(0)}$ to be identical.  
With the same eigenfunctions as the gauge bosons,  the free Lagrangian for the scalar fields are given by 
\bea
\mathcal{L}_4^H &\supset& 
 - \frac{1}{2} \left( \frac{1}{g} \right)^2 h^{(0)}\left( \Box_{4} + m_h^2 \right) h^{(0)} 
- \frac{1}{2} \left( \frac{1}{\sqrt{2} \, g} \right)^2  \sum_{n=1}^{\infty} 
\left[ h^{(n)}
  \left( \Box_{4}+ \left(m_h^2 + m_{n}^2 \right) \right) h^{(n)} \right] \nonumber \\
 &&+ \left\{h^{(m)} \to \phi^{(m)}, \; m_h \to 0\right\}
\eea 
The U(1) gauge symmetry is broken by $\langle H \rangle =v/\sqrt{2}$, and the U(1) gauge boson acquires its mass. 
After normalizing the kinetic terms for all zero-modes and KK-modes, 
    we find the gauge boson masses,
\bea 
   m_A^{(0)} = Q_H  g v_h,  \quad m_A^{(n)} = \sqrt{m_n^2 + (Q_H g v_h)^2}, 
\eea
 for the zero-mode and the KK-modes, respectively. 
Here we have defined $v_h$ as $v_h =v/g$ by considering the normalizations.

\subsection{Domain-wall fermions} 
Next is the consideration of 5-dimensional fermions. 
Here we also consider the U(1) gauge theory to simply our discussion, 
  which can be easily extended to the 5-dimensional SM case.  
We follow a mechanism in Ref.~\cite{rubakov}
 to generate the domain-wall fermion in 5-dimensional space-time.

Let us first consider a real scalar field ($\varphi(x, y)$) in  the 5-dimensional bulk: 
\begin{eqnarray} 
 {\cal L}_{(5)} = 
  \frac{1}{2} \left(\partial_{M} \varphi \right) \left( \partial^{M} \varphi \right) - V(\varphi)   \; ,  
\end{eqnarray}
where the scalar potential is give by 
\bea 
 V(\varphi) = \frac{m_\varphi^4}{2 \lambda} 
	   -m_\varphi^2 \varphi^2  
           + \frac{\lambda}{2} \varphi^4 .  
\eea
We may assign $Z_2$-odd parity for $\varphi$. 
This parity assignment is consistent with having a non-trivial solution to the equation of motion, 
 namely, the so-called kink solution,  
\bea
\varphi_0 (y) = \frac{m_\varphi}{\sqrt{\lambda}} 
\tanh [m_\varphi y] .  
\label{kink} 
\eea

Following Ref.~\cite{rubakov}, we introduce the Lagrangian for a bulk fermion coupling with $\varphi$, 
\bea
\mathcal{L}&=&i \overline{\psi}\left[\gamma^{\mu}D_{\mu}+i\gamma^{5}D_{y}\right]\psi+Y \varphi \overline{\psi}\psi \nonumber \\ 
&=& 
i \overline{\psi_L} \gamma^{\mu}D_{\mu} \psi_{L}+i \overline{\psi_R} \gamma^{\mu}D_{\mu} \psi_{R}  \nonumber \\
&-& \overline{\psi_L} D_{y}\psi_{R}+\bar{\psi}_{R}D_{y} \psi_L + Y \varphi \left( \overline{\psi_L}\psi_{R}+ \overline{\psi_R}\psi_{L}\right), 
\eea
where we decompose the Dirac fermion $\psi$ into its chiral components, $\psi= \psi_L + \psi_R$, 
  the covariant derivative is given by $D_M=\partial_M - i Q_f A_M$ with a U(1) charge $Q_f$ for $\psi$, 
  and $Y$ is a positive constant.   
We define $Z_2$-parity for $\psi_L$ and $\psi_R$ to be even and odd, respectively. 
Neglecting the gauge interaction and replacing $\varphi$ by the kink solution, 
  the equations of motion are given by 
\bea
&& i\gamma^{\mu} \partial_{\mu}\psi_{L}- \partial_{y}\psi_{R} + Y \varphi_0 \psi_{R}=0,  \nonumber \\
&& i\gamma^{\mu}  \partial_{\mu}\psi_{R}+ \partial_{y}\psi_{L} + Y \varphi_0 \psi_{L}=0.
\eea
Since we are interested in the zero-mode solution that corresponds to a SM chiral fermion, 
   we focus on the equations of motion for zero-modes: 
   $\psi_L (x,y) \supset \psi_L^{(0)}(x) \, \chi_L^{(0)}(y)$  and  $\psi_R (x,y) \supset \psi_R^{(0)}(x) \, \chi_R^{(0)}(y)$, 
   such that\footnote{
See Ref.~\cite{HO} for complete analysis. 
}    
\bea
\left(\frac{d}{dy}+Y \varphi_0 \right) \chi^{(0)}_{L} = 0,  \quad 
\left(\frac{d}{dy} -Y  \varphi_0 \right) \chi^{(0)}_{R} =0. 
\eea
In the vicinity of $y=0$, $\varphi_0 \simeq m_\varphi^2 y/\sqrt{\lambda}$, and we can find the approximate solutions as 
\bea
  \chi_{L}^{(0)} = C_L e^{- \frac{M_F^2}{2} y^2}, \quad 
  \chi_{R}^{(0)} = C_R e^{+ \frac{M_F^2}{2} y^2}, 
\eea 
where $C_{L,R}$ are constants, and $M_F=\sqrt{Y m_\varphi^2/\sqrt{\lambda}}$. 
Since $\chi_R$ is $Z_2$-odd, $C_R=0$  and the right-handed chiral fermion is projected out.\footnote{
Here, $Z_2$-parity is not essential. Even without introducing the $Z_2$-parity, 
  the right-handed chiral fermion is delocalizing at $y=0$ and thus it is not normalizable \cite{rubakov}.  
}
Therefore, we have only one chiral fermion in the 4-dimensional effective theory. 
We fix $C_L=\sqrt{M_F/\sqrt{\pi}}$ and canonically normalize the fermion kinetic term.

Now we describe the Lagrangian for the chiral fermion in the 4-dimensional effective theory as 
\bea
 {\mathcal L}_4 &\supset&  \left[ \int_{-\infty}^{\infty}  dy \left( \chi_{L}^{(0)} \right)^2 \right]  
  \overline{\psi}_L^{(0)} i \gamma^\mu \left( \partial_\mu - i Q_f A_\mu^{(0)} \right) \psi_L^{(0)} 
 \nonumber \\ 
&+& 
\sum_{n=1}^\infty  Q_f \left[ \int_{-\infty}^{\infty} dy  \left( \chi_{L}^{(0)} \right)^2  \cos(m_n y) \right] 
 A_\mu^{(n)}  \left[\overline{\psi}_L^{(0)} \gamma^\mu \psi_L^{(0)} \right]
\eea
Rescaling the gauge fields to canonically normalize their kinetic terms in Eq.~(\ref{4DA}), 
  we obtain the final expression as 
\bea
 {\mathcal L}_4 &\supset&  
  \overline{\psi}_L^{(0)} i \gamma^\mu \left( \partial_\mu - i Q_f g A_\mu^{(0)} \right) \psi_L^{(0)} 
+ \sum_{n=1}^\infty  Q_f g_{\rm eff}^{(n)} A_\mu^{(n)}  \left[\overline{\psi}_L^{(0)} \gamma^\mu \psi_L^{(0)} \right], 
\eea
where the 4-dimensional effective coupling between the chiral fermion and the KK-mode gauge boson is given by  
\bea
   g_{\rm eff}^{(n)} = \sqrt{2} \, g \, \exp\left[- \frac{1}{4} \left( \frac{m_n}{M_F} \right)^2 \right]. 
\eea
The widths of localized gauge fields and the chiral fermion are characterized by $1/m_n$ and $1/M_F$, respectively. 
A limit $M_F \to \infty$ corresponds to the case that the chiral fermion is localized on a ``3-brane" with zero-width
  and all KK-modes of the bulk gauge boson universally couple with the fermion.  
On the other hand, $g_{\rm eff}^{(n)}$ is vanishing in the limit of $M_F \to 0$. 
This is analogous to the case in the universal extra-dimension model \cite{UED}, 
  where such couplings are forbidden by the momentum conservation in the 5th coordinate direction.

\begin{figure}[t]
\begin{center}
\includegraphics[scale=1.5]{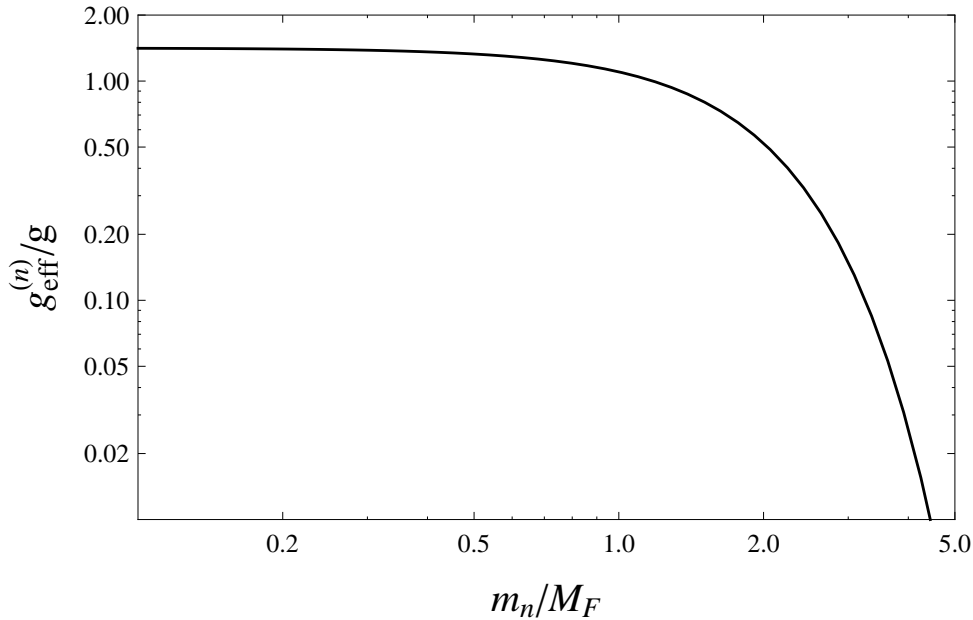}       
\end{center}
\caption{
The effective coupling ($g_{\rm eff}^{(n)}$) between the $n$-th KK-mode gauge boson and the chiral fermion 
  as a function of $m_n/M_F$. 
The gauge coupling of the zero-mode gauge boson is denoted as $g$.   
}
\label{fig:1}
\end{figure}

In Figure \ref{fig:1}, we show the effective gauge coupling between the $n$-th KK-mode gauge boson 
  and the chiral fermion.   
For a narrow width of the domain-wall fermion, the effective gauge coupling approaches to $\sqrt{2} \, g$. 
As the width of the domain-wall fermion becomes large, the effective coupling is decreasing exponentially.  
When applied to the SM, the gauge coupling $g$ corresponds to one of the SM gauge couplings 
  and the chiral fermion is identified with an SM fermion.  
We will discuss implications of this coupling behavior to the LHC phenomenology in the next section.

In 5 dimensions, we may introduce the Yukawa coupling of the SM fermions as 
\bea
  {\mathcal L}_Y = - Y_f \overline{D} H S +{\rm H.c.} 
  =- Y_f \overline{D_L} H S_R  - Y_f \overline{D_R} H S_L + {\rm H.c.}, 
\eea
where $D$ and $S$ are Dirac fermions in 5 dimensions, 
  whose chiral components, $D_L$ and $S_R$, are identified 
  as left-handed SM doublet and right-handed singlet fermions under the SM SU(2) gauge group, respectively, 
  by assigning them $Z_2$-even parities. 
With the couplings for $D$ and $S$ with the kink background, 
  zero-modes of $D_L$ and $S_R$ are localized around $y=0$, 
  while zero-modes of $D_R$ and $S_L$ are projected out.  
The Higgs Lagrangian in Eq.~(\ref{L_H}) is now extended to the SM Higgs doublet case. 
If we simply set the same couplings for $D$ and $S$ with the kink background, 
  in other words, the common widths of the domain-wall fermions,  
  we obtain an effective Yukawa coupling as the same as $Y_f$ in the 4-dimensional effective theory.

\section{LHC Phenomenology}
\label{sec:3}

The ATLAS and the CMS collaborations have been searching for a new gauge boson resonance 
  with a variety of final states at the LHC Run-2.  
For the sequential SM $Z^\prime$ and $W^\prime$ bosons, 
  which have the same properties as the SM $Z$ and $W$ bosons except for their masses, 
  the ATLAS collaboration has recently reported their search results with luminosity of about 36 fb$^{-1}$. 
The lower bound on the sequential SM $Z^\prime$ boson is obtained to be $m_{Z^\prime} \geq 4.5$ TeV 
  with dilepton final states \cite{ATLAS:2017},   
  while $m_{W^\prime} \geq 5.1$ TeV for the sequential SM $W^\prime$ boson mass is obtained 
  with its decay mode $W^\prime \to l \nu$  \cite{ATLAS2:2017}.    
In the following, we interpret these results as constraints on the KK-mode gauge bosons in our scenario.

For simplicity, we set a common width $2 L$ for the $y$-dependent SM gauge couplings, 
  so that the KK-mode mass spectra for gluon, weak bosons and photon are approximately the same 
  for $m_{W,Z}^2 \ll m_1^2=(\pi/L)^2$ (this relation will be justified below).   
Thus, let us consider the most severe constraint from the $W^\prime$ boson search. 
Since the total decay width of $W^\prime$ boson is about 3\% of its mass for $m_{W^\prime} \gtrsim 1$ TeV, 
   we employ the narrow-width approximation in evaluating the parton-level cross section of the process, 
\bea 
   \hat{\sigma}(q \overline{q}^\prime \to W^\prime) \propto \Gamma_{W^\prime}(W^\prime \to q \overline{q}^\prime)  
      \, \delta(M_{\rm inv}^2 -m_{W^\prime}^2)  \propto  g^2,  
\eea   
where $M_{\rm inv}^2$ is the invariant mass of the initial partons, 
   $\Gamma_{W^\prime}(W^\prime \to q \overline{q}^\prime)$ is the partial decay width into $q \overline{q}^\prime$, 
   and $g$ is the SM SU(2) gauge coupling. 
When we identify the $W^\prime$ boson with the 1st KK-mode of the SM $W$ boson in the Domain-Wall SM,  
   the only difference is from the effective gauge coupling $g_{\rm eff}^{(1)}$. 
As shown in Figure \ref{fig:1}, the effective coupling $g_{\rm eff}^{(1)}$ is a function of the width of the domain-wall fermion. 
For simplicity, we assume a common width for all SM fermions.    
Hence, in the narrow-width approximation, we have 
\bea
   \sigma (pp \to W^{(1)} \to l \nu) = \left( \frac{g_{\rm eff}^{(1)}}{g} \right)^2 \sigma (pp \to W^\prime \to l \nu),  
\eea
  by which we can interpret the current ATLAS constraints as those on the 1st KK-mode $W$ boson. 

\begin{figure}[h]
\begin{center}
\includegraphics[scale=1.3]{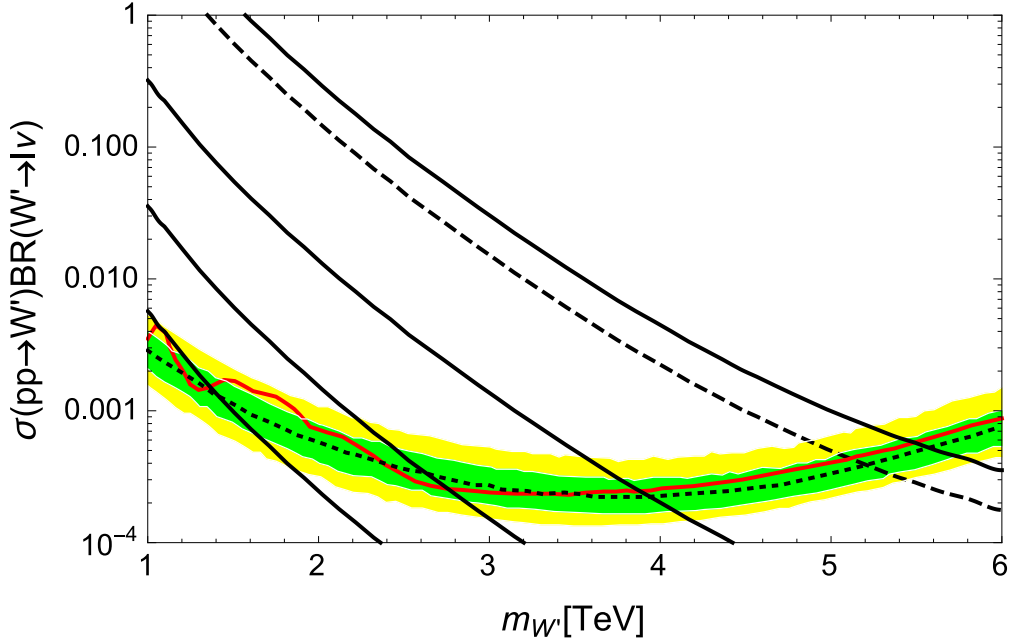}       
\end{center}
\caption{
The cross section $\sigma (pp \to W^{(1)} \to l \nu)$ as a function of $m_{W^\prime}=m_1$
  for $g_{\rm eff}^{(1)}/g=0.04$, $0.1$, $0.3$ and $\sqrt{2}$ (solid diagonal lines) from left to right, 
  along with the theoretical prediction of $\sigma (pp \to W^\prime \to l \nu)$ for the sequential SM $W^\prime$ boson (dashed diagonal line) 
  and the ATLAS results \cite{ATLAS2:2017} 
  (the cross section upper bound (solid horizontal curve (in red)) at 95\% Confidence Level,  
   the expected limit (dotted horizontal curve), $\pm 1 \sigma$ expectation (green shaded) 
  and $\pm 2 \sigma$ expectation (yellow shaded)).}
\label{fig:2}
\end{figure}

In Figure \ref{fig:2}, we show the cross section $\sigma (pp \to W^{(1)} \to l \nu)$ as a function of $m_{W^\prime}=m_1$
  for various values of $g_{\rm eff}^{(1)}/g$, 
  along with the upper bound on the cross section from the ATLAS results \cite{ATLAS2:2017} 
  at the LHC RUn-2 with a 36.1 fb$^{-1}$ integrated luminosity (horizontal solid curve (in red))
  and the theoretical prediction of $\sigma (pp \to W^\prime \to l \nu)$ for the sequential SM $W^\prime$ boson (dashed line). 
The solid diagonal lines from left to right depict the theoretical predictions 
  of the cross section $\sigma (pp \to W^{(1)} \to l \nu)$ as a function of $m_1$
  for $g_{\rm eff}^{(1)}/g=0.04$, $0.1$, $0.3$ and $\sqrt{2}$, respectively. 
We see from Figure \ref{fig:1} that these $g_{\rm eff}^{(1)}/g$ values correspond to $m_1/M_F=3.8$, $3.3$, $2.5$ and $0$, 
  respectively.  
For a fixed $g_{\rm eff}^{(1)}/g$ value, we read off the lower bound on the 1st KK-mode mass 
  from the intersection of the corresponding solid diagonal line and the solid horizontal (red) curve.  
We find the lower bounds on the 1st KK-mode mass as 
  $m_1({\rm TeV}) \geq 1.4$, $2.8$, $3.9$ and $5.5$ 
  for $g_{\rm eff}^{(1)}/g=0.04$, $0.1$, $0.3$ and $\sqrt{2}$, respectively.

Let us compare this current LHC bound with constraints imposed by the electroweak precision test (EWPT) measurements. 
In Ref.~\cite{Davoudiasl:1999tf}, the authors have considered the effective four Fermi operators induced by the bulk SM gauge bosons in the Randall-Sundrum (RS) model and have obtained a lower bound on the 1st KK-mode mass as $m_1({\rm TeV}) \geq 23$. 
In the model the KK-mode coupling is enhanced as $g_{\rm eff}^{(1)}/g\simeq 8.4$. 
Considering, for example, $g_{\rm eff}^{(1)}/g\simeq \sqrt{2}$ in our model, we can easily interpret the lower bound as $m_1({\rm TeV}) \geq 23\times \left(\sqrt{2}/8.4\right) \simeq 3.8$. 
Hence, for this coupling, the current LHC bound is more severe. 
This is also true for a general choice of   $g_{\rm eff}^{(1)}/g$. 
As another constraint from the EWPT measurements, we consider contributions to the $S$ and $T$ parameters \cite{Peskin:1991sw} from the KK-modes loop corrections. 
In our model, the mass spectrum of the gauge and the Higgs boson KK-modes, as well as their couplings to the SM gauge bosons are the same as those in the Universal Extra Dimension (UED) model \cite{Appelquist:2000nn}. 
In Ref.~\cite{Baak:2011ze}, the Gfitter Group has studied the $S$ and $T$ parameter constraints for the UED model and obtained  
a lower bound $m_{1}({\rm TeV}) \gtrsim 1$  (at 2-$\sigma$ confidence level). 
Hence, we can see from Fig.~\ref{fig:1} that the LHC constraints are more severe than the $S$ and $T$ parameter constraints for $g_{\rm eff}^{(1)}/g \gtrsim 0.04$.

\section{Conclusions and discussions}
\label{sec:4}

We have proposed a framework to construct ``Domain-Wall SM" which is defined in a non-compact 5-dimensional space-time. 
Considering localization mechanisms for the gauge field, the Higgs field and the chiral fermion 
  in the 5-dimensional Minkowski space,  
  we have derived the 4-dimensional effective Lagrangian for the SM fields 
  and the gauge boson KK-modes.  
The effective gauge couplings between the KK-modes and the SM chiral fermions 
  are controlled by their domain-wall widths. 
This geometrical property provides us with an interesting LHC phenomenology on the KK-modes of the SM gauge bosons.  
We have interpreted the current LHC results from the search for a new gauge boson resonance 
  as the constraints on the Domain-Wall SM.

In the present paper, we have introduced $Z_2$-parity under the reflection of the 5th coordinate. 
This is only for simplifying our formulas and not essential for the construction of the Domain-Wall SM. 
In the absence of $Z_2$-parity, we can consider the case that the SM chiral fermions are localized 
  around different points in the 5th dimension. 
Such a generalization opens up a possibility to solve the fermion mass hierarchy problem in the SM from the wave-function overlapping, leading to an exponentially suppressed effective Yukawa couplings, 
  as proposed in Ref.~\cite{AS}. 
In general, such a setup can potentially generate flavor-dependent 
KK-mode gauge boson couplings and hence the flavor changing neutral currents (FCNCs)  mediated by the KK-gauge bosons. 
It is worth investigating in detail a setup, where fermions are localized in different positions along the extra dimension direction to naturally reproduce the fermion mass hierarchy while avoiding the FCNC constraints. We leave these considerations for our future work. 
Configurations of the domain-wall fermions reflect their effective gauge couplings with the KK-mode gauge bosons. 
Therefore, this ``geometry" in the 5th dimension can be tested at the future  LHC experiment,  
  once a KK-mode gauge boson is discovered and its coupling to the SM fermions is measured. 
We may also consider an extension of the Domain-Wall SM, for example, the grand unified theory in non-compact extra-dimensions. 
There is an interesting proposal in Ref.~\cite{Eto_et_al} for a possibility to break the grand unified gauge group into the SM one 
  via domain-wall configurations. 
Hence, the Domain-Wall SM can provide us with a variety of interesting phenomenologies.

Since graviton resides in the bulk, we also need to consider a localization of graviton 
  to complete our proposal of the Domain-Wall SM. 
For this purpose, we may combine our scenario with the RS-2 scenario \cite{RS2} 
  with the Planck brane at $y=0$. 
Here we may identify the Planck brane as a domain-wall with the zero-width limit. 
The mass spectrum of the KK-modes of the SM fields is controlled 
  by the width of the domain-walls, and the current LHC results constrain it to be $\lesssim$(1 TeV)$^{-1}$. 
On the other hand, the width of 4-dimensional graviton is controlled by the AdS curvature $\kappa$ 
   in the RS-2 scenario and its experimental constraint is quite weak, $\kappa \gtrsim 10^{-3}$ eV \cite{RS2}. 
Therefore, we can take $\kappa \ll 1$ TeV and neglect the warped background geometry 
  in our setup of the Domain-Wall SM, while the 4-dimensional Einstein gravity is reproduced in the RS-2 scenario 
  at low energies. 
We may think if the energy density from the SM domain-walls is large and affects the RS-2 background geometry. 
However, we expect the energy density from the domain-walls of ${\cal O}(\Lambda^4)$ with $\Lambda={\cal O}$(1 TeV), 
  while the energy density of the Planck brane in the RS-2 scenario is given by ${\cal O}(M_P^2 \kappa^2)$ 
  with the reduced Planck mass of $M_P\simeq 2.4 \times10^{18}$ GeV.  
Therefore, we choose the AdS curvature in the range of $10^{-3}$ eV$\ll \kappa \ll$1 TeV for the theoretical consistency of our scenario.

\section*{Acknowledgements}
The authors would like to thank Minoru Eto and Masato Arai for useful discussions. 
This work of N.O. is supported in part by the U.S. Department of Energy (DE-SC0012447).



\end{document}